\documentclass[12pt]{article}

\ifx\pdfoutput\undefined
\usepackage[dvips,bookmarks]{hyperref}
\else
\usepackage{hyperref}
\fi
\hypersetup{colorlinks=false,bookmarksopen,bookmarksnumbered,citecolor=blue,
   pdfstartview=FitH}

\usepackage{latexsym}

\usepackage{amssymb,amsfonts,amsmath}
\usepackage{graphicx} 
\usepackage{indentfirst}

 \usepackage{bbm}

\parskip=\medskipamount

\allowdisplaybreaks

\newcommand {\cB}{{\cal B}}

\newcommand {\cD}{{\cal D}}

\newcommand {\cN}{{\cal N}}

\def\a{\alpha}

\def\b{\beta}
\def\c{\chi}
\def\d{\delta}
\def\e{\epsilon}

\def\g{\gamma}

\def\k{\kappa}
\def\l{\lambda}

\def\L{\Lambda}
\def\O{\Omega}

\def\S{\Sigma}

\def\rd{{\rm d}}
\def\ri{{\rm i}}
\def\re{{\rm e}}

\newcommand{\ad}{{\dot{\alpha}}}                           
\newcommand{\bd}{{\dot{\beta}}}                          

\newcommand{\be}{\begin{equation}}
\newcommand{\ee}{\end{equation}}
\newcommand{\bea}{\begin{eqnarray}}
\newcommand{\eea}{\end{eqnarray}}

\newcommand{\ba}{\begin{array}}
\newcommand{\ea}{\end{array}}

\def\double #1{#1{\hbox{\kern-2pt $#1$}}}

\newcommand{\gd}{{\dot\g}}
\newcommand{\dd}{{\dot\d}}

\newcommand{\bsubeq}{\begin{subequations}}
\newcommand{\esubeq}{\end{subequations}}

\begin{document}

\begin{titlepage}

\begin{center}
{\Large \bf  Goldstino superfields in $AdS_4$}
\end{center}

\begin{center}

{\large  
I. N. McArthur\footnote{{ian.mcarthur@uwa.edu.au}}
} \\
\vspace{5mm}

\footnotesize{
{\it School of Physics, M013, 
The University of Western Australia\\
35 Stirling Highway, Crawley W.A. 6009 \\
Australia }}  
~\\

\vspace{2mm}

\end{center}
\vspace{5mm}

\begin{abstract}
\baselineskip=14pt
Spontaneous breaking of global supersymmetries results in Goldstino fields which provide a nonlinear realisation of the supersymmetry algebra. A finite supersymmetry transformation of a Goldstino field can be used to generate a superfield whose components provide a linear realisation of the supersymmetry algebra.  This construction also automatically determines the action of the algebra of supercovariant derivatives on Goldstino superfields, essential to the efficient computation  of manifestly supersymmetric component actions for the Goldstinos, including coupling to matter fields. In this paper, we extend known constructions of Goldstino superfields resulting from spontaneous breaking of supersymmetry in flat four-dimensional $\cN =1$ superspace to spontaneous breaking of  $\cN =1$ and $\cN =2$ supersymmetry in $AdS_4. $\end{abstract}
\vspace{1cm}

\vfill
\end{titlepage}

\tableofcontents{}
\vspace{1cm}
\bigskip\hrule

\section{Introduction}
\setcounter{equation}{0}

Supersymmetry is clearly not realised in nature as an unbroken symmetry, as evidenced by the absence of  equal mass superpartners for known elementary particles. There are nevertheless compelling theoretical grounds for the realisation of supersymmetry as a broken symmetry, including a natural solution to the ``hierarchy problem''.

Sparked by the work of Komargodski and Seiberg 
\cite{Komargodski:2009rz}, there has  recently been a revival of interest in techniques by which a Goldstino, providing a nonlinear realisation of supersymmetry, can be promoted to a superfield \cite{Ivanov:1977my, IK1978,IK1982,Uematsu:1981rj,Samuel:1982uh}. This allows the use of standard superfield techniques for  the construction of supersymmetric actions for the Goldstino, including coupling to matter multiplets.  

Given a  realization of the supersymmetry algebra on a set of fields (usually including auxiliary fields), a corresponding superfield can be constructed via a finite supersymmetry transformation of the field of lowest mass dimension in the supermultiplet with the supersymmetry transformation parameter $\epsilon$ replaced by the fermionic coordinate $\theta$ of superspace;
  for a review see \cite{WB}. The corresponding superfield codes a linear realization of supersymmetry on its component fields. 

Spontaneous breaking of global ${\cal N}=1$ supersymmetry gives rise to a nonlinear realization of the supersymmetry algebra on a Goldstino field $\l_{\a}(x)$ \cite{Volkov:1972jx,Akulov:1974xz}. In the spirit of the previous paragraph, a finite supersymmetry transformation of the Goldstino with parameter $\theta$ gives rise to a ``Goldstino superfield'' $\L_{\a}(x, \theta)$ whose lowest component is $\l_{\a}(x),$ and whose  higher components are expressed in terms of the Goldstino $\l_{\a}(x)$ and its derivatives. By construction, the components of this superfield provide a linear realization of the supersymmetry algebra \cite{Ivanov:1977my, IK1978,IK1982,Uematsu:1981rj,Samuel:1982uh}. 

Of crucial significance is the action of the algebra of supercovariant derivatives on the the Goldstino superfield, as it is this algebra that allows the components of the superfield,  and component expressions for corresponding actions, to be calculated efficiently.  In the case of spontaneous breaking of ${\cal N}= 1$ supersymmetry in flat superspace, Samuel and Wess \cite{Samuel:1982uh} provided a construction to determine the action of the algebra super covariant derivatives on the Goldstino superfield. 

However, their construction requires modification in the case where the underlying superspace is curved, and this paper considers the case of such coset superspaces. We illustrate the modified construction in  the case of the breaking of ${\cal N}= 1$ and ${\cal N}= 2$ global supersymmetry in four-dimensional anti-de Sitter space ($AdS_4$), but it is applicable to any coset superspace.

 Keck provided the first treatment of ${\cal N}= 1$ supersymmetry in four-dimensional anti-de Sitter space $AdS_4,$ and Ivanov and Sorin extensively developed the corresponding superfield techniques \cite{Ivanov:1980vb}. 
 Deser and Zumino undertook the first analysis of nonlinearly realized ${\cal N}= 1$ supersymmetry in $AdS_4$  \cite{Deser:1977uq}, and the results were subsequently recast in the standard language of nonlinear realizations \cite{Coleman:1969sm,Callan:1969sn,Isham:1969ci} by Zumino \cite{Zumino:1977av}, providing a systematic derivation of the nonlinear transformations and the action for the resulting Goldstino $\l_{\a}(x).$ The techniques presented in this paper provide a more modern perspective on these results in terms of Goldstino superfields. We also consider the case of spontaneously broken $\cN = 2$ supersymmetry in $AdS_4.$
 
 The outline of the paper is as follows. In section 2, we review standard results on nonlinear realisations of broken symmetries. The construction of Goldstino superfields from Goldstino fields and the method for determining the realisation of the algebra of supercovariant derivatives on these superfields is discussed in section 3. These results are applied to the case of spontaneous breaking of $\cN =1$ supersymmetry in $AdS_4$ in section 4 using the so-called ``chiral'' coset parameterisation for the Goldstino fields. Section 5 treats the same case using the Volkov-Akulov coset parameterisation. The paper concludes with some results for spontaneous breaking of $\cN =2$ supersymmetry in $AdS_4.$

\section{Nonlinear transformations of Goldstone fields}
\setcounter{equation}{0}

Before proceeding to an analysis of Golstino superfields in curved coset superspaces more general than Minkowski superspace, we will review the general techniques  for determining the nonlinear transformations of the Goldstone fields  that result when a  symmetry is spontaneously broken \cite{Coleman:1969sm,Callan:1969sn,Isham:1969ci}. They are applicable to spontaneously broken bosonic and fermionic symmetries.

 Consider a global symmetry group $G$ with generators $\{T_i , T_a\},$ where the $T_i$ generate spontaneously broken symmetries, and the $T_a$ generate a subgroup $H$ of unbroken symmetries. The Goldstone bosons $\xi^i$ corresponding to the broken symmetries paramaterize the coset $G/H$ via a `slice' through $G$ of the form 
\be
 g(\xi) = \re^{\ri \, \xi^i T_i}.
 \label{slice}
 \ee
The realization of $G$ on the Goldstone fields is determined by
$$ g. g(\xi) =  g(\xi') \, h(g, \xi),$$
where $ h(g, \xi) \in H$ is a compensating transformation to return to the slice (\ref{slice}) \cite{Coleman:1969sm,Callan:1969sn,Isham:1969ci}.

The action of group elements  $g = \re^{\ri \e^i T_i}$ corresponding to the broken symmetries  is nonlinear on the Goldstone fields. We want to determine $\d \xi^i = \xi'^i - \xi^i$ when $\e$ is infinitesimal. In infinitesimal form,
$$ (1 + \ri \, \e^i T_i) \, g(\xi) =  g(\xi + \d \xi) \, (1+ \ri \, H(\xi, \e)).$$
Writing $ g(\xi + \d \xi) = g(\xi) + \d g(\xi),$ 
$$  \ri \, \e^i T_i \, g(\xi) =  \delta g(\xi ) + \ri \, g(\xi) \, H(\xi, \e), $$
which implies
\be  \ri \,g(\xi)^{-1} \, \e^i T_i \, g(\xi) =  g(\xi)^{-1} \delta g(\xi ) +  \ri \, H(\xi, \e ). \label{Zumino} 
\ee
Equations determining $\d \xi^i$ are obtained equating the coefficients of $T_i$ on both sides of this equation \cite{Zumino:1977av}.

Using Zumino's notation \cite{Zumino:1977av}
$$X \wedge Y = [X,Y], \quad X^2 \wedge Y = [X, [X,Y]],$$
 the left and right hand sides of (\ref{Zumino}) are obtained via
$$  g(\xi)^{-1} \, \e^i T_i \, g(\xi)  = \re^{- \ri \, \xi^jT_j} \wedge  \e^i T_i , \quad 
g(\xi)^{-1} \delta g(\xi ) = \frac{(\re^{- \ri \, \xi^j T_j} - 1)}{- \ri \, \xi^kT_k} \wedge \ri \, \delta \xi^i T_i. $$

\section{``Bridging'' the supersymmetry algebra and the algebra of supercovariant derivatives}
\setcounter{equation}{0}

Spontaneous breaking of a global ${\cal N} =1$ supersymmetry gives rise to a Goldstino field on which the supersymmetry algebra is realized nonlinearly  \cite{Volkov:1972jx,Akulov:1974xz}. The Goldstino field will be denoted generically by $\l_{\a}(x).$  In this section, we review how to construct a superfield $\L_{\a}( x, \theta, \bar{\theta})$ whose components are composites of the Goldstino and its derivatives, and how to determine the action of the algebra of supercovariant derivatives on this superfield. The construction generalises the flat space construction of \cite{Samuel:1982uh}, in that it is applicable to general coset superspaces.

Consider a superalgebra that allows the construction of a superspace as a coset. For example,  $\cN =1$ $AdS_4$ superspace can be realised as the coset $OSp(1|4)/SO(3,1).$  Let $Q_A = (P_a, Q_{\a}, \bar{Q}^{\ad})$ denote the generators of superspace translations in the super algebra,  and $M_I$ denote the Lorentz generators and, in the case of extended supersymmetry, R-symmetry generators. 
Introducing superspace coordinates $Z^A = (x^a, \theta^{\a}, \bar{\theta}_{\ad}),$ the coset representative of a point in superspace can be chosen to be
$$ g(Z) = \re^{\ri Z^A Q_A}.$$
The Maurer-Cartan form
\be g(Z)^{-1} d  g(Z) =  \ri \, E^A (Z) Q_A + \ri \, \O^I (Z) \, M_I 
\ee
yields the supervielbein $E^A(Z)$ on superspace and  a connection $\O^I(Z)$ associated with the Lorentz subgroup.
Using the decompositions
$ d =  E^A(Z) \, D_A , \, \O^I (Z) =  E^A (Z) \, \O_A{}^I (Z), $ the Maurer-Cartan equation can be written as
\be D_A \, g(Z) =  \ri \, g(Z) \, Q_A + \ri \,g(Z)\,  \O_A{}^I (Z) \, M_I .
\label{Maurer-Cartan}
\ee
Defining the covariant derivative
 $ \cD_A \, g(Z)  = D_A \, g(Z) - \ri \, g(Z) \, \O_A{}^I (Z) \,M_I,$ then equation (\ref{Maurer-Cartan}) can be written
\be \cD_A \, g(Z) = \ri \, g(Z) \, Q_A.
\label{bridge}
\ee
Similar techniques have been used  in the specific context of $\cN = 1$ supersymmetry in $AdS_5$  \cite{Bagger:2011na}, related to earlier work in 
\cite{Kuzenko:2001ag}, \cite{Kuzenko:2007aj}.

The significance of equation (\ref{bridge}) is the following. We can construct a superfield from the nonlinear realization of the supersymmetry algebra on the Goldstino fields via
\be 
\L_{\a} (Z) = g(Z) \times \l_{\a} (0),
\ee
where the cross is the notation of Wess and Bagger \cite{WB} to denote the  action of the superspace translation generators $Q_A = (P_{\a \ad}, Q_{\a}, \bar{Q}_{\ad})$ on the Goldstone fields.  If we apply both sides of equation (\ref{bridge}) to $\l_{\a}(0),$ then we obtain
\be
\cD_A \, \L_{\a} (Z) = \ri \, g(Z) \, Q_A \times \l_{\a} (0).
\label{curved}
\ee
In other words,  $g(Z)$ can be considered as providing a ``bridge'' between realizations of the algebra of the superspace translations $Q_A$ on a space of Goldstone fields $\l_{\a} (0),$ {\it evaluated at the  origin of superspace,} and the algebra of supercovariant derivatives acting on the superfields $ \L_{\a} (Z).$

The above construction provides a  coset superspace generalization of the original method used by Samuel and Wess \cite{Samuel:1982uh} in Minkowski superspace to determine the algebra of supercovariant derivatives on Goldstino superfields. Since Samuel and Wess were working in flat superspace, they constructed Goldstino superfields in the form
\be 
\L_{\a} (Z) = e^{ \ri (\theta^{\a} Q_{\a} + \bar{\theta}_{\ad} \bar{Q}^{\ad})} \times \l_{\a} (x),
\ee
where the spatial dependence of the Goldstino superfield was determined from the spatial dependence of the Goldstino $\l_{\a}(x).$ In curved superspace, this is no longer appropriate, as the generators $P_{\a \ad}$ of spatial translations in general do not commute with the supersymmetry generators $Q_{\a}.$  We will also see shortly that only requiring knowledge of the realization of the generators $Q_{A}$ on $\l_{\a}$ at the origin and not at a general spatial point leads to considerable simplifications in the methods outlined in section 2.

In the following sections, we will develop the ``bridge'' (\ref{curved}) in the case of spontaneously broken $\cN=1$  and $\cN=2$ supersymmetry in $AdS_4.$  

\section{Spontaneously broken ${\cal N} = 1$ supersymmetry in $AdS_4$}
\setcounter{equation}{0}

In 1969, systematic coset constructions  of nonlinearly realized {\it internal} symmetries were developed  \cite{Coleman:1969sm, Callan:1969sn, Isham:1969ci}. Volkov \cite{Volkov} extended the coset construction to include both broken and unbroken {\it spacetime} symmetries (see also \cite{Og}), and Volkov and Akulov \cite{Volkov:1972jx,  Akulov:1974xz}  applied it to the case of broken supersymmetries.

In the context of  flat $\cN = 1$ superspace in four space-time dimensions, the Volkov-Akulov nonlinear realization of supersymmetry on a Goldstino field $\l_{\a}(x)$ is based on the group element
\be
g\big(x, \l (x), \bar{\l} (x)\big) = {\rm e}^{\ri( - x^a P_a + \l^{\a}(x) Q_{\a} + \bar{\l}_{\ad}(x) \bar{Q}^{\ad})}~.
\label{VA}
\ee
Rigid supersymmetry transformations of the Goldstino are generated via  left action by the group element
\be
g (\e, \bar{\e}) =  {\rm e}^{\ri(\e^{\a} Q_{\a} + \bar{\e}_{\ad} \bar{Q}^{\ad})}
\label{ge}
\ee
on $g\big(x, \l (x), \bar{\l} (x)\big).$ Supersymmetry transformations mix $\l_{\a}$ and $\bar{\l}_{\ad}.$

Zumino  \cite{Zumino:ChiralNLSusy} introduced an alternative ``chiral'' nonlinear realization, that was further
developed by Samuel and Wess \cite{Samuel:1982uh}. It involves a Goldstino $\xi_{\a}(x)$ 
which mixes only with itself (and not with $\bar{\xi}_{\ad}$) under supersymmetry transformations.
It was shown in \cite{KM} that this  nonlinear realization is related to the alternative coset parametrization
\be
g\big(x, \xi (x), \bar{\psi} (x)\big) 
= {\rm e}^{\ri( - x^a P_a + \xi^{\a}(x) Q_{\a})} \, {\rm e}^{\ri  \bar{\psi}_{\ad}(x) \bar{Q}^{\ad}}~.
\label{Z}
\ee
Supersymmetry transformations of $\bar{\psi}_{\ad}$ do mix $\bar{\psi}_{\ad}$ and $\xi_{\a}$.
In treating breaking of ${\cal N}=1$ supersymmetry in $AdS_4,$ we will begin with this ``chiral'' nonlinear realization of the superalgebra on Goldstinos.

Four-dimensional anti de Sitter space $AdS_4$ can be realized as the homogeneous space $SO(3,2)/SO(3,1).$ In an appropriate basis,  the generators of $SO(3,2)$ consist of the generators  $M_{\a \b}$ of the Lorentz subgroup $SO(3,1)$ supplemented by translation generators $P_{\a \ad}$:
\bea
\, [ M_{\a \b}, M_{\g \d} ] &=& - \, \frac{\ri}{2} \, \left(\e_{\a \g} M_{\b \d} + \e_{\b \g} M_{\a \d} + \e_{\a \d} M_{\b \g} + \e_{\b \d}M_{\a \g} \right) \\
\, [ \bar{M}_{\ad \bd}, \bar{M}_{\gd \dd} ] &=& - \, \frac{\ri}{2} \, \left(\e_{\ad \gd} \bar{M}_{\bd \dd} + \e_{\bd \gd} \bar{M}_{\ad \dd} + \e_{\ad \dd} \bar{M}_{\bd \gd} + \e_{\bd \dd}\bar{M}_{\ad \gd} \right) \\
\, [ M_{\a \b}, P_{\g \gd}] &=& - \, \frac{\ri}{2} \, \left( \e_{\a \g} P_{\b \gd} + \e_{\b \g} P_{\a \gd} \right) \\
\, [ \bar{M}_{\ad \bd}, P_{\g \gd}] &=& - \, \frac{\ri}{2} \, \left( \e_{\ad \gd} P_{\g \bd} + \e_{\bd \gd} P_{\g \ad} \right) \\
\, [ P_{\a \ad}, P_{\b \bd} ] &=& -\, \frac{\ri}{2} \, |\k|^2 \, \left( \e_{\a \b} \bar{M}_{\ad \bd} + \e_{\ad \bd} M_{\a \b} \right).
\eea
The complex parameter $\kappa$ has dimensions of mass; the scalar curvature of $AdS_4$ is determined by $|\kappa|.$

The  ${\cal N} = 1$ supersymmetric extension of $AdS_4$ is the coset superspace $OSp(1|4)/SO(3,1),$ where the generators of $OSp(1|4)$ include those of $SO(3,2)$  appended by supersymmetry generators $Q_{\a}$ obeying the algebra:

\bea
\, [ M_{\a \b} , Q_{\g}] &=& - \, \frac{\ri}{2} \, \left( \e_{\a \g} Q_{\b} + \e_{\b \g} Q_{\a}\right) \\
\, [ \bar{M}_{\ad \bd} , \bar{Q}_{\gd }] &=& - \, \frac{\ri}{2} \, \left( \e_{\ad \gd}  \bar{Q}_{\bd } + \e_{\bd \gd}  \bar{Q}_{\ad } \right) \\
\, \{ Q_{\a}, \bar{Q}_{\ad } \} &=& 2 \,  P_{\a \ad} \\
\, \{ Q_{\a}, Q_{\b} \} &=& 2\, \k \, M_{\a \b}  \\
\, \{ \bar{Q}_{\ad }, \bar{Q}_{\bd } \} &=& 2\, \bar{ \k} \, \bar{M}_{\ad \bd} \\
\, [ P_{\a \ad}, Q_{\b}] &=& - \, \frac{\ri}{2} \, \k \, \e_{\a \b}\,   \bar{Q}_{\ad } \\
\, [ P_{\a \ad}, \bar{Q}_{\bd }  ] &=& - \, \frac{\ri}{2} \,\bar{\k} \, \e_{\ad \bd}  \,  Q_{\a}
\eea
The phase associated with the complex parameter $\kappa$ can be absorbed into a redefinition of $Q_{\a}$ and $\bar{Q}_{\ad}.$

As emphasized in section 3, to determine the action of supercovariant derivatives on the superfields corresponding to the Goldstinos, we only need to know the transformation of the Goldstino fields at the origin i.e. we only need to compute 
\be
 \re^{ \ri Y} \, g(0,  \xi(0), \bar{\psi}(0)) 
 \label{Yaction}
 \ee
where $Y = - a^a P_a + \e^{\a} Q_{\a} - \bar{\e}^{\ad} Q_{\ad}.$ This simplifies the computation considerably. Now,
$$ \re^{ \ri Y} \, g(0,  \xi(0), \bar{\psi}(0)) = g(x', \xi'(x'), \bar{\psi}'(x')) \, \re^{\ri H} $$
where $H$ generates a compensating Lorentz transformation. If we apply the formula (\ref{Zumino}) to the infinitesimal version of this expression, the variations $\d \xi$ and $ \d \bar{\psi}$ it yields are of the form
$$ \d \xi = \xi'(\d x) - \xi(0), \quad \d \bar{\psi} = \bar{\psi}'( \d x) - \bar{\psi} (0).$$
The supersymmetry transformations require comparison of the fields $\xi$ and $\xi'$ at the same point, in this case the origin, and similarly for $\bar{\psi}.$ This introduces an additional shift so that the infinitesimal supersymmetry transformations take the form
\be
 \tilde{\d} \xi (0) = \delta \xi - \d x^a \partial_a \xi(0), \quad  \tilde{\d} \bar{\psi} (0) = \delta \bar{\psi} - \d x^a \partial_a \bar{\psi} (0).
 \label{SStransf}
 \ee

Using the ${\cal N}=1$ superymmetry algebra $OSp(1|4),$ and with fields $\xi^{\a}$ and $\bar{\psi}^{\ad}$ evaluated at the origin, the method for determining nonlinear transformations based on equation (\ref{Zumino}) yields the following:

\noindent
Coefficient of $Q_{\a}$:
\bea
& & ( 1 - \frac{3 \ri}{4} \, \k \, \xi^2 ) ( 1 + \frac{\ri}{2} \,\bar{ \k }\bar{\psi}^2 ) \, \e^{\a} - \ri \, \bar{\k} \,  \xi^{\a} \bar{\psi}_{\ad} \, \bar{\e}^{\ad} - \frac14 \bar{\k} (1+ \frac{\ri}{4} \k \xi^2) \, a^{\ad \a} \bar{\psi}_{\ad} \nonumber  \\ &=&( 1 - \frac{\ri}{4} \, \k \, \xi^2)  ( 1 + \frac{\ri}{2} \, \bar{\k} \, \bar{\psi}^2 ) \, \d \xi^{\a} - \, \frac{1}{4} \, \bar{\k} \, (1 + \frac{\ri}{12} \, \k \, \xi^2 )  \bar{\psi}_{\ad} \, \d x^{\ad \a}.
\eea

\noindent
Coefficient of $\bar{Q}_{\ad}$:
\bea
& & - \, (1 - \frac{\ri}{2} \, \k \, \xi^2) (1 + \frac{ 3 \ri}{4} \, \bar{\k} \, \bar{\psi}^2)  \, \bar{\e}^{\ad} + \frac14 \k (1 + \frac{3 \ri}{4} \bar{\k} \bar{\psi}^2) \, a^{\ad \a} \xi_{\a} \nonumber \\
& = & \frac18 \, \k \, (1+ \frac{3 \ri}{4} \, \bar{\k} \bar{\psi}^2) \xi_{\a} \, \d x^{\ad \a} - (1 + \frac{\ri}{4} \, \bar{\k} \, \bar{\psi}^2) \, \d \bar{\psi}^{\ad} .
\eea

\noindent
Coefficient of $P_{\a \ad}$:
\bea
& & - \, 2 \ri \, (1 - \frac{3 \ri}{4} \, \k \, \xi^2) \bar{\psi}^{\ad} \, \e^{\a} - 2 \ri \, ( 1 - \frac{\ri}{4} \, \bar{\k } \, \bar{\psi}^2) \, \xi^{\a} \, \bar{\e}^{\ad} + \frac12 (1+ \frac{\ri}{4} \k \xi^2) \, ( 1 - \frac{\ri}{4} \bar{\k} \bar{\psi}^2) \, a^{\ad \a} \nonumber \\
& = & - \, 2 \ri \, (1 - \frac{\ri}{4} \, \k \, \xi^2) \bar{\psi}^{\ad} \, \d \xi^{\a} + \frac12 \, (1 + \frac{\ri}{12} \, \k \, \xi^2 ) (1 - \frac{\ri}{4} \, \bar{\k} \, \bar{\psi}^2 ) \, \d x^{\ad \a}.
\eea


These equations can be solved to give
\bea
\d x^{\ad \a} &=& - \, 4 \ri \, \xi^{\a} \bar{\e}^{\ad} + (1 + \frac{\ri}{6} \k \, \xi^2) \, a^{\ad \a}\\
\delta \xi^{\a} &=& (1 - \frac{\ri}{2} \, \k \, \xi^2 ) \, \e^{\a} \\
\d \bar{\psi}^{\ad} &=& ( 1 + \frac{\ri}{2} \, \bar{\k} \, \bar{\psi}^2) \, \bar{\e}^{\ad} + \frac14 \k (1 + \frac{\ri}{3} \k \, \xi^2) \, (1+ \frac{\ri}{2} \bar{\k} \,\bar{\psi}^2) \, \xi_{\a} a^{\ad \a} .
 \eea
Inserting these results into equation (\ref{SStransf}), the infinitesimal supersymmetry transformations for the Goldstone fields at the origin are
\bea
\tilde{\d} \xi^{\a}  &=& (1 + \frac{\ri}{2} \, \k \, \xi^2) \, \e^{\a} - 2 \ri \, \xi^{\b} \bar{\e}^{\bd} \partial_{\b \bd} \xi^{\a}  \\
\tilde{\d} \bar{\psi}^{\ad}  &=&  ( 1 + \frac{\ri}{2} \, \bar{\k} \, \bar{\psi}^2) \, \bar{\e}^{\ad} - 2 \ri \, \xi^{\b} \bar{\e}^{\bd} \partial_{\b \bd} \bar{\psi}^{\ad} .
\eea
Using the fact that $\tilde{\d} \xi  = \ri  Y \times \xi,$ as follows from equation (\ref{Yaction}), and similarly for $\bar{\psi},$ it follows that
\bea
\ri  P_{\a \ad} \times \xi^{\b} &=& (1 + \frac{\ri}{6} \k \, \xi^2) \, \partial_{\a \ad} \xi^{\b}  \label{Pxi} \\
\ri  Q_{\a} \times \xi^{\b}  &=& \d_{\a}{}^{\b} \,  (1 - \frac{\ri}{2} \, \k \, \xi^2) \label{Qxi} \\
\ri  \bar{Q}_{\ad}  \times \xi^{\b}  &=&  - 2 \ri \, \xi^{\a} \partial_{\a \ad} \xi^{\b}  \\
\ri  P_{\a \ad} \times \bar{\psi}^{\bd} &=& \frac12 \k \,\d_{\ad}{}^{\bd}  (1+ \frac{\ri}{3} \k \, \xi^2) \, ( 1 + \frac{\ri}{2} \bar{\k} \, \bar{\psi}^2) \, \xi_{\a}  + (1+\frac{\ri}{6} \k \, \xi^2) \, \partial_{\a \ad} \bar{\psi}^{\bd} \label{Ppsi} \\
\ri  Q_{\a} \times \bar{\psi}^{\bd}  &=& 0 \\
\ri  \bar{Q}_{\ad}  \times \bar{\psi}^{\bd}  &=& - \, \d_{\ad}{}^{\bd} \,  ( 1 + \frac{\ri}{2} \, \bar{\k} \, \bar{\psi}^2)  -  2 \ri \, \xi^{\a}  \partial_{\a \ad} \bar{\psi}^{\bd}  \label{barQpsi}
\eea
where all fields are evaluated at the origin.

As shown in section 3, the `bridge' between the realization of the algebra of supercovariant derivatives on superfields and the nonlinear realization of the superspace generators on the space of fields $\xi^{\a}(0)$ and $\bar{\psi}^{\ad}(0)$ is
$$ g(Z) = \re^{\ri \,(   x^a P_a + \theta^{\a}Q_{\a} - \bar{\theta}^{\ad}\bar{Q}_{\ad})}, $$
in the sense that
$$ \cD_a \, g(Z) =  g(Z) \, \ri P_a, \quad \cD_{\a} \, g(Z) = g(Z) \, \ri Q_{\a}, \quad \bar{\cD}_{\ad} \, g(Z) = g(Z) \, \ri \bar{Q}_{\ad} . $$
This allows us to map the nonlinear realization of  the superalgebra  on the fields $\xi^{\a} (0)$ and $\bar{\psi}(0)$ to a realization of the algebra of supercovariant derivatives on the superfields
$$ \Xi^{\a} (Z) = g(Z) \times \xi^{\a} (0), \quad \bar{ \Psi}^{\ad}(Z) = g(Z) \times \bar{\psi}^{\ad} (0). $$
From (\ref{Pxi}) and (\ref{Ppsi}) and the bridging relation  $ \cD_a \, g(Z) =  g(Z) \, \ri P_a,$ we obtain 
\bea
 g(Z) \times \partial_{\a \ad} \, \xi^{\b}(0) &=&  (1 - \frac{\ri}{6} \k \, \Xi^2) \, \cD_{\a \ad} \,  \Xi^{\b} \\
 g(Z) \times \partial_{\a \ad} \, \bar{\psi}^{\bd} &=&  (1 - \frac{\ri}{6} \k \, \Xi^2) \,\cD_{\a \ad} \, \bar{\Psi}^{\bd} - \frac12 \k \,  \d_{\ad}{}^{\bd} ( 1 + \frac{\ri}{2} \bar{\k} \, \bar{\Psi}^2) \, \Xi_{\a} .
 \eea
The remaining relations in (\ref{Qxi}) - (\ref{barQpsi}) yield
\bea
\cD_{\a} \Xi^{\b} &=&  \d_{\a}{}^{\b} \,  (1 - \frac{\ri}{2} \, \k \, \Xi^2) \label{D1} \\
\bar{\cD}_{\ad} \Xi^{\b} &=&  - 2 \, \ri \, \Xi^{\a} \,  \cD_{\a \ad} \, \Xi^{\b} \label{D2}  \\
\cD_{\a} \bar{\Psi}^{\bd} &=& 0 \\
\bar{\cD}_{\ad} \bar{\Psi}^{\bd} &=& - \, \d_{\ad}{}^{\bd} \, (1 + \frac{\ri}{2} \k \, \Xi^2 ) \, ( 1 + \frac{\ri}{2} \, \bar{\k} \, \bar{\Psi}^2)  -  2\, \ri \, \Xi^{\a}  \, \cD_{\a \ad} \, \bar{\Psi}^{\bd} . 
\eea

Noting that
$$ M_{\a \b} \times \L_{\g} = \frac{\ri}{2} \, (\e_{\a \g} \L_{\b} + \e_{\b \g} \L_{\a} ), $$
it can be verified by explicit calculation that this provides a realization of the algebra of covariant derivatives\footnote{Comparing, for example with equation (5.5.6) in \cite{Buchbinder:1998qv}, this corresponds to a supergravity background with $R = - \frac{\ri}{2} \bar{\k},$ $G_{\a \ad} = 0,$ $W_{\a \b \gamma} = 0,$ and the Lorentz generators are scaled by a factor $ \ri.$}
\bea
\, \{ \cD_{\a}, \bar{\cD}_{\ad} \} &=& - 2 \, \ri\ \cD_{\a \ad} \nonumber \\
\, [ \cD_{\a \ad} , \cD_{\b} ] &=& - \, \frac12 \, \k \, \e_{\a \b} \, \bar{\cD}_{\ad}  \nonumber \\
\, [ \cD_{\a \ad} ,\bar{ \cD}_{\bd} ] &=& - \, \frac12 \, \bar{ \k} \, \e_{\ad \bd} \, \cD_{\a}  \nonumber  \\
\, \{ \cD_{\a} , \cD_{\b} \} &=&  2 \k \, M_{\a \b}  \nonumber  \\
\, \{ \bar{ \cD}_{\ad} , \bar{ \cD}_{\bd} \} &=&  2 \bar{\k} \, \bar{ M}_{\ad \bd}  \nonumber  \\
 \, [ { \cD}_{\a \ad} , { \cD}_{\b \bd} ] &=& - \, \frac{\ri}{2} \, |\k|^2  \left( \e_{\a \b} \bar{M}_{\ad \bd} + \e_{\ad \bd} M_{\a \b} \right). \label{coderivs}
\eea

It should be noted that Samuel and Wess \cite{Samuel:1982uh}, ``after some guesswork'', arrived at a realization of the algebra of supercovariant derivatives on the superfields $\Xi_{\a}$ in a general curved superspace of the form
\bea
\, \cD_{\a} \Xi^{\b} &=&  \d_{\a}{}^{\b} \,  (1  - R^* \, \Xi^2) \label{SW1}\\
\bar{\cD}_{\ad} \Xi^{\b} &=& - 2 \ri \, \Xi^{\a} \cD_{\a \ad} \Xi^{\b} + \frac12  \,\Xi^2 G_{\ad}{}^{\b}, \label{SW2}
\eea
where the superfields $R$ and $G_{\a \ad}$ are curved superspace analogues of the Ricci scalar and the Einstein tensor.\footnote{We have set the parameter $k$ determining the scale of supersymmetry breaking in  \cite{Samuel:1982uh} to 1. We also use the conventions of \cite{Buchbinder:1998qv} in which the chiral projector is $(\bar{\cD}^2 - 4R).$}
The results (\ref{D1}) and (\ref{D2}) are consistent with (\ref{SW1}) and (\ref{SW2}) via the identifications  $G_{\a \ad} = 0$ and $R^* = \frac{i}{2} \k.$ 
In fact

Kuzenko and Tyler  \cite{KT} provided justification for the ``guess'' (\ref{SW1}) and (\ref{SW2}) by expressing $\Xi_{\a}$ as
\be
 \Xi_{\a} = \frac12 \cD_{\a} \bar{\S}
\ee
where $\bar{\S}$ is a constrained complex linear superfield. In the case of $AdS_4,$  we have an explicit realisation of $\bar{\S}$ as
\be
\bar{\S} = \Xi^{\a} \Xi_{\a},
\ee
which satisfies the nilpotency condition $\bar{\S}^2 = 0$ and the constraints
\be
- \frac14 (\cD^2 - 2 \ri \k ) \, \bar{\S} = 1, \quad
\bar{\S} \, \cD^2 \bar{\cD}_{\ad} \bar{\S} = \bar{\cD}_{\ad} \bar{\S}
\ee
as a result of (\ref{D1}) and (\ref{D2}).
This generalises a flat superspace construction  given in \cite{KM}. 

Similarly, the covariantly chiral superfield
\be
\Phi = \Psi^{\a} \Psi_{\a}
\ee
satisfies the nilpotency condition  $\Phi^2 = 0$ and the constraint
\be
- \, \frac14 \Phi \, (\bar{\cD}^2 + 2 \ri \bar{\k}) \bar{\Phi} = \Phi,
\ee
generalising a flat superspace construction by Rocek \cite{Rocek}. 

The covariantly chiral superfield $\Phi$ allows for construction of an action (using the notation and conventions of \cite{Buchbinder:1998qv})
\bea
S & = &  - \frac12 \, \int \rd^8 z \, \frac{E^{-1}}{\bar{R}} \, \bar{\Phi} + c.c \\
&=&  - \frac12 \,  \int \rd^4x \, e^{-1} \, (- \frac14 \bar{\cD}^2 + B) \, \bar{\Phi}| + c.c., 
\eea
where $|$ denotes the $\theta = 0 = \bar{\theta}$ projection of a superfield and $B = 3 \bar{R}|.$
The algebra of fermionic covariant derivatives on the superfields $\Xi^{\a}$ and $\bar{\Psi}_{\ad},$ it is possible to express $\bar{\cD}^2 \, \bar{\Phi}$ in terms of bosonic covariant derivatives acting on 
$\Xi^{\a}$ and $\bar{\Psi}_{\ad}.$ Using $\Xi^{\a}(x, \theta, \bar{\theta})| = \xi^{\a}(x)$ and $ \bar{\Psi}_{\ad}(x, \theta, \bar{\theta})| = \bar{\psi}_{\ad} (x), $ and defining $ \cD_{\a \ad} \, \Xi | = D_{\a \ad} \, \xi $ and $ \cD_{\a \ad} \, \bar{\Psi}| = D_{\a \ad} \, \bar{\psi} ,$ 
\bea
S = &-&   \int \rd^4 x \,e^{-1} \, (\,  \frac12 + \ri \, \xi^{\a} D_{\a \ad} \, \bar{\psi}^{\ad} + \frac{ \ri }{2} \k \, \xi^2 - \frac{ \ri }{2} \bar{\k} \, \bar{\psi}^2
  - \, \xi^{\a}  (D_{\a \ad} \, \xi_{\b}) D^{\ad \b} \, \bar{\psi}^2 \nonumber \\
  &- & \frac14 \, \xi^2 D^{\ad \a}D_{\a \ad} \, \psi^2  - \frac{1}{2} \k \, \xi^2 \, \bar{\psi}^{\ad} D_{\a \ad} \,\xi^{\a}  - \frac{1}{4} |\k|^2 \, \xi^2 \, \bar{\psi}^2 ) + c.c.
  \label{SWaction}
 \eea
 In particular, there is  a  mass term  for the Goldstino with mass inversely proportional to the radius of anti de Sitter space, consistent with the results in \cite{Deser:1977uq,Zumino:1977av}. As in the flat  case \cite{KM}, whilst this action is apparently at most fourth order in Goldstinos, the fields $ (\xi_{\a}, \bar{\xi}^{\ad})$ and $ (\psi_{\a}, \bar{\psi}^{\ad})$ are related by nonlinear transformations via (\ref{VA}) and (\ref{Z}); expressing the action in terms of the fields $ (\xi_{\a}, \bar{\xi}^{\ad})$ alone  or  $ (\psi_{\a}, \bar{\psi}^{\ad})$ alone will result in terms to eighth order in Goldstino fields, in accord with the general analysis in \cite{KT}.

\section{The Volkov-Akulov realization in $\cN = 1$ $AdS_4$}
As discussed in the previous section, the Volkov-Akulov nonlinear realization of the supersymmetry algebra on Goldstinos (as opposed to the ``chiral'' representation of Zumino and Samuel and Wess)  is based on a coset parameterization
$$ g(Z) = \re^{ \ri \, ( -x^aP_a + \l^{\a}(x) Q_{\a} + \bar{\l}_{\ad} (x) Q^{\ad})}.$$
We now repeat the steps in the previous section for this parameterization.

With fields evaluated at $x=0,$
\bea
\d x^{\ad \a} &=& - \, 2 \ri \, (1 + \frac{\ri}{8} \, \k \, \l^2) \, \bar{\l}^{\ad} \e^{\a} - 2  \ri \, ( 1 - \frac{\ri}{8} \bar{\k} \bar{ \l}^2) \, \l^{\a} \bar{\e}^{\ad} \nonumber  \\
& & + (1 + \frac{\ri}{24} \, \k \, \l^2 - \frac{\ri}{24} \, \bar{\k} \, \bar{\l}^2 + \frac{17}{1440} |\k |^2 \l^2 \bar{\l}^2) \, a^{\ad \a}\\
\d \l^{\a} &=& (1 - \frac{\ri}{2} \, \k \, \l^2 + \frac{\ri }{12} \, \bar{\k} \, \bar{\l}^2 - \frac{1}{288} \, |\k|^2 \, \l^2 \, \bar{\l}^2 ) \, \e^{\a} - \, \frac{\ri}{12} \, \bar{\k} \l^{\a} \bar{\l}_{\ad} \bar{\e}^{\ad} \nonumber \\
& &  - \, \frac18 \, \bar{\k} \,( 1 - \frac{\ri}{8} \, \k \, \l^2 ) \, \bar{\l}_{\ad} \, a^{\ad \a}\\
\d \bar{\l}^{\ad} &=& (1 - \frac{\ri}{12} \k \l^2 + \frac{\ri}{2} \bar{\k} \bar{\l}^2 - \frac{13}{288} |\k|^2 \l^2 \bar{\l}^2 ) \, \bar{\e}^{\ad}  - \, \frac{\ri}{12} \, \k \, \bar{\l}^{\ad} \l_{\a} \e^{\a}  \nonumber \\
& &  - \frac18 \, \k \, (1 + \frac{\ri}{8} \, \bar{\k} \, \bar{\l}^2 ) \, \l_{\a} \, a^{\ad \a}.
\eea

Recalling that supersymmetry transformations are of the form
$ \tilde{\d} \l (0) = \delta \l - \d x^a \partial_a \l (0)$ with
$$ \tilde{\d} = \ri \, (- a^a P_a + \e^{\a} Q_{\a} - \bar{\e}^{\ad} \bar{Q}_{\ad}) \times, $$
we obtain
\bea
\ri  P_{\a \ad} \times \l^{\b} & = &  ( 1 + \frac{\ri}{24} \k \l^2 - \frac{\ri}{24} \bar{\k} \bar{\l}^2 + \frac{17}{1440} | \k|^2 \l^2 \bar{\l}^2 ) \, \partial_{\a \ad} \l^{\b} \label{Pl}\\
& &  - \, \frac14 \bar{\k} \, \delta_{\a}{}^{\b} (1 - \frac{\ri}{8} \k \l^2) \, \bar{\l}_{\ad}   \\
\ri Q_{\a}\times \l^{\b} &=& \d_{\a}{}^{\b} \, (1 - \frac{\ri}{2} \k \, \l^2 + \frac{\ri}{12} \bar{\k} \, \bar{\l}^2 - \frac{1}{288} | \k|^2 \l^2 \bar{\l}^2) \nonumber \\
& & + \ri \, (1 + \frac{\ri}{8} \k \, \l^2) \, \bar{\l}^{\ad} \partial_{\a \ad} \l^{\b} \label{51} \\
\ri \bar{Q}_{\ad} \times \l^{\b} &=& \frac{\ri}{12} \, \bar{\k} \, \l^{\b} \bar{\l}_{\ad} - \, \ri \, (1 - \frac{\ri}{8} \bar{\k} \, \bar{\l}^2) \, \l^{\a} \partial_{\a \ad} \l^{\b} \label{52} \\
\ri  P_{\a \ad} \times \bar{ \l}^{\bd} & = &  ( 1 + \frac{\ri}{24} \k \l^2 - \frac{\ri}{24} \bar{\k} \bar{\l}^2 + \frac{17}{1440} | \k|^2 \l^2 \bar{\l}^2 ) \, \partial_{\a \ad} \bar{\l}^{\bd} \nonumber \\
& &  - \, \frac14 \k \, \delta_{\ad}{}^{\bd} (1 + \frac{\ri}{8} \bar{\k} \bar{\l}^2) \, \l_{\a}  \label{Pbarl} \\
\ri Q_{\a} \times \bar{\l}^{\bd} &=&  \frac{\ri}{12} \, \k \, \l_{\a}  \bar{\l}^{\bd} + \ri \, (1 + \frac{\ri}{8} \k \, \l^2) \, \bar{\l}^{\ad} \partial_{\a \ad} \bar{ \l}^{\bd} \label{54}\\
\ri \bar{Q}_{\ad} \times \bar{\l}^{\bd} &=& - \, \d_{\ad}{}^{\bd} \, (1 - \frac{\ri}{12} \k \, \l^2 + \frac{\ri}{2} \bar{\k} \, \bar{\l}^2 - \frac{1}{288} | \k|^2 \l^2 \bar{\l}^2) \nonumber \\
& &  - \, \ri \, (1 - \frac{\ri}{8} \bar{\k} \, \bar{\l}^2) \, \l^{\a} \partial_{\a \ad} \bar{ \l}^{\bd}. \label{55}
\eea

The Goldstino superfields are constructed as 
$$ \L^{\a} (Z) = g(Z) \times \l^{\a} (0), \quad \bar{ \L}^{\ad}(Z) = g(Z) \times \bar{\l}^{\ad} (0). $$
From (\ref{Pl}) and (\ref{Pbarl}) and the `bridging' relation 
 $ \cD_a \, g(Z) =  g(Z) \, \ri P_a,$ we obtain 
\bea
 g(Z) \times \partial_{\a \ad} \, \l^{\b}(0) &=& ( 1 - \frac{\ri}{24} \k \l^2 + \frac{\ri}{24} \bar{\k} \bar{\l}^2 - \frac{1}{120} | \k|^2 \l^2 \bar{\l}^2 ) \, \cD_{\a \ad} \L^{\b} \nonumber \\
 & & + \frac14 \bar{\k} \, \delta_{\a}{}^{\b} (1 - \frac{\ri}{6} \k \L^2) \, \bar{\L}_{\ad}  \\
 g(Z) \times \partial_{\a \ad} \,\bar{\l}^{\bd}(0) &=& ( 1 - \frac{\ri}{24} \k \l^2 + \frac{\ri}{24} \bar{\k} \bar{\l}^2 - \frac{1}{120} | \k|^2 \l^2 \bar{\l}^2 ) \, \cD_{\a \ad} \bar{\L}^{\bd} \nonumber \\
& & + \frac14 \k \, \delta_{\ad}{}^{\bd} (1 + \frac{\ri}{6} \bar{\k} \bar{\L}^2) \, \L_{\a} 
 \eea
 Using this and the remaining the `bridging' relations 
 $$ \cD_{\a} \, g(Z) = g(Z) \, \ri Q_{\a}, \quad \bar{\cD}_{\ad} \, g(Z) = g(Z) \, \ri \bar{Q}_{\ad} ,$$
 we obtain obtain from (\ref{51}), (\ref{52}), (\ref{54}), (\ref{55})
 \bea
 \cD_{\a} \L^{\b} &=& \d_{\a}{}^{\b}  ( 1 - \frac{\ri}{2} \k \L^2 - \frac{\ri}{6} \bar{\k} \bar{\L}^2 - \frac{1}{72} | \k|^2   \L^2 \bar{\L}^2 ) \nonumber \\
 & & + \, \ri \, (1 + \frac{\ri}{12} \k \L^2) \, \bar{\L}^{\ad} \cD_{\a \ad} \L^{\b} \\
\bar{\cD}_{\ad} \L^{\b} &=& - \, \ri \, ( 1 - \frac{\ri}{12} \bar{\k} \bar{\L}^2 ) \, \L^{\a} \cD_{\a \ad} \L^{\b} + \frac{\ri}{6} \, \bar{\k} \bar{\L}_{\ad} \L^{\b} \\
\cD_{\a} \bar{\L}^{\bd} &=&  \ri \, ( 1 + \frac{\ri}{12} \k \L^2) \, \bar{\L}^{\ad} \cD_{\a \ad} \bar{\L}^{\bd} - \frac{\ri}{6} \, \k \L_{\a} \bar{\L}^{\bd} \\
\bar{\cD}_{\ad}  \bar{\L}^{\bd} &=& - \, \d_{\ad}{}^{\bd} ( 1 + \frac{\ri}{6} \k \L^2 + \frac{\ri}{2} \bar{\k} \bar{\L}^2 - \frac{1}{72} | \k|^2 \L^2 \bar{\L}^2) \nonumber \\
& & - \, \ri \,  ( 1 - \frac{\ri}{12} \bar{\k} \bar{\L}^2 ) \, \L^{\a} \cD_{\a \ad} \bar{\L}^{\bd}.
\eea
Again a highly nontrivial check on these results is that the algebra of covariant derivatives closes in the form (\ref{coderivs}).

A natural action for the Goldstino is (using the conventions and notation of \cite{Buchbinder:1998qv})
\bea
S &=& \int \rd^8 z \, E^{-1} \, \L^2 \bar{\L}^2 \nonumber \\
&=&- \, \frac14  \int \rd^8 z \, \frac{E^{-1}}{R} (\bar{\cD}^2 - 4 R) \, \L^2 \bar{\L}^2 \nonumber \\
&=&  \int \rd^8 z \, \frac{E^{-1}}{R} \c
\eea
with 
\be
 \c= \L^2 + \ri \, \L^2 \bar{\L}^{\ad} \cD_{\a \ad} \L^{\ad} - \frac{ 2 \ri}{3} \bar{\k} \L^2 \bar{\L}^2 + O(\L^6).
\ee
Using 
\be 
\int \rd^8 z \, \frac{E^{-1}}{R} \c = \int \rd^4 x \, e^{-1} (- \frac14 \cD^2 + B) \, \c |
\ee
with $B = 3 \bar{R}),$
the action to quadratic order in the Goldstinos is
\be
S = \int \rd^4 x \, e^{-1} \, \left( 1 + \ri \bar{\l}^{\ad} D_{\a \ad} \l^{\a} + \ri \l^{\a} D_{\a \ad} \bar{\l}^{\ad} + \ri \k \l^2 - \ri \bar{\k} \bar{\l}^2 \right).
\label{VAaction}
\ee
The Akulov-Volkov Goldstinos  $(\l_{a}, \bar{\l}^{\ad})$ are related to the Samuel-Wess Goldstinos $(\xi_{\a}, \bar{\psi}^{\ad})$ by nonlinear transformations via (\ref{VA}) and (\ref{Z}). To leading order, $\xi_{\a} = \l_{\a} + \cdots,$ $ \bar{\psi}^{\ad} = \bar{\l}^{\ad} + \cdots,$ which means the kinetic and mass terms in the actions (\ref{SWaction}) and (\ref{VAaction}) are equivalent.

\section{Broken ${\cal N} = 2$ supersymmetry in $AdS_4$}

The  ${\cal N} = 2$ supersymmetric extension of $AdS_4$ is the coset superspace $OSp(2|4)/SO(3,1)$ ${\rm x} \, SO(2),$ where $SO(3,1)$ is the Lorentz subgroup. The algebra contains two supercharges $Q_{\a}^i, \, i = 1,2$ with $\bar{Q}_{\ad i} = (Q_{\a}^i)^*,$ and $SU(2)$ generators $J_{ij}$ mixing the two supercharges.  In addition to the complex parameter $\k$ determining the curvature of $AdS_4,$ the algebra contains a constant real iso-triplet $S_{ij}$  which determines the torsion of ${\cal N} = 2$ superspace in $AdS_4$ \cite{Kuzenko:2008ep, Kuzenko:2008qw}.\footnote{ $S_{ij} = S{ji}= \epsilon_{ik}\, \epsilon_{jl}S^{kl},$  $S^2 = \frac12 \, S^{ij}S_{ij}.$} We choose a basis for the generators of the algebra $OSp(2|4)$  that yields the algebra 
\setcounter{equation}{0}
\bea
\, [ M_{\a \b}, M_{\g \d} ] &=& - \frac{\ri}{2} \, \left(\e_{\a \g} M_{\b \d} + \e_{\b \g} M_{\a \g} + \e_{\a \d} M_{\b \g} + \e_{\b \d}M_{\a \g} \right) \\
\, [ \bar{M}_{\ad \bd}, \bar{M}_{\gd \dd} ] &=& - \frac{\ri}{2} \, \left(\e_{\ad \gd} \bar{M}_{\bd \dd} + \e_{\bd \gd} \bar{M}_{\ad \gd} + \e_{\ad \dd} \bar{M}_{\bd \gd} + \e_{\bd \dd}\bar{M}_{\ad \gd} \right) \\
\, [ M_{\a \b}, P_{\g \dd}] &=& - \frac{\ri}{2} \, \left( \e_{\a \g} P_{\b \gd} + \e_{\b \g} P_{\a \gd} \right) \\
\, [ \bar{M}_{\ad \bd}, P_{\g \dd}] &=& - \frac{\ri}{2} \, \left( \e_{\ad \gd} P_{\b \bd} + \e_{\bd \gd} P_{\g \ad} \right) \\
\, [ M_{\a \b} , Q_{\g}^i ] &=& - \frac{\ri}{2} \, \left( \e_{\a \g} Q_{\b}^i + \e_{\b \g} Q_{\a}^i \right) \\
\, [ \bar{M}_{\ad \bd} , \bar{Q}_{\gd i}] &=& - \frac{\ri}{2} \, \left( \e_{\ad \gd}  \bar{Q}_{\bd i} + \e_{\bd \gd}  \bar{Q}_{\ad i} \right) \\
\, [ J_{ij}, Q_{\a}^k] &=& - \frac{\ri}{2} \, \left(\delta_i{}^k Q_{\a j} + \delta_j{}^k Q_{\a i} \right) \\
\, [ J_{ij}, \bar{Q}_{\ad k} ] &=&  \frac{\ri}{2} \, \left(\e_{ik}  \bar{Q}_{\ad j}  + \e_{jk}  \bar{Q}_{\ad i}  \right) \\
\, \{ Q_{\a}^i, \bar{Q}_{\ad j} \} &=& 2 \, \delta^i{}_j \, P_{\a \ad} \\
\, \{ Q_{\a}^i, Q_{\b}^j \} &=& 2\, \k \, S^{ij} M_{\a \b} + \k \, \e_{\a \b} \, \e^{ij} \,S^{kl}J_{kl} \\
\, \{ \bar{Q}_{\ad i}, \bar{Q}_{\bd j} \} &=& 2\, \bar{ \k} \, S_{ij} \bar{M}_{\ad \bd} - \bar{ \k}\, \e_{\ad \bd} \, \e_{ij} \, S^{kl}J_{kl} \\
\, [ P_{\a \ad}, Q_{\b}^i ] &=& - \frac{\ri}{2} \, \k \, \e_{\a \b}\,  S^{ij} \bar{Q}_{\ad j} \\
\, [ P_{\a \ad}, \bar{Q}_{\bd i}  ] &=& - \frac{\ri}{2} \,\bar{\k} \, \e_{\ad \bd}  \, S_{ij} Q_{\a}^j \\
\, [ P_{\a \ad}, P_{\b \bd} ] &=& - \frac{\ri}{2} \, |\k|^2 S^2 \, \left( \e_{\a \b} \bar{M}_{\ad \bd} + \e_{\ad \bd} M_{\a \b} \right).
\eea

We consider the Samuel-Wess coset parameterization of the coset,
\be
g = e^{\, \ri \, (- x^a P_a + \xi^{\a}_i (x) Q^i_{\a})} \, e^{i \bar{\psi}_{\ad}^i (x) \bar{Q}_i^{\ad} }.
\label{SWN=2}
\ee
As in the case of flat superspace and $\cN = 1$ superspace in $AdS_4,$  the Goldstinos $\xi^{\a}_i (x) $ mix only with themselves under supersymmetry transformations (and not with the $\bar{\psi}_{\ad}^i (x)$). The reason is that when acting with $e^{\ri \, (\epsilon^{\a}_i  Q^i_{\a}+\bar{\epsilon}_{\ad}^i \bar{Q}_i^{\ad})}$ from the left on $g$ as parameterised in (\ref{SWN=2}), terms generated in the first exponential $e^{\, \ri \, (- x^a P_a + \xi^{\a}_i (x) Q^i_{\a})}$ are proportional to $P, Q, \bar{Q}, M$ and $S.J.$ The terms proportional to $P$ and $Q$ contribute to the supersymmetry transformations of $x$ and $\xi.$ The remaining terms must be ``moved to the right'' into and through the second exponential factor $e^{ \ri \bar{\psi}_{\ad}^i (x) \bar{Q}_i^{\ad} }.$ In the process, no terms proportional to $Q$ are produced that could feed back into a $\bar{\psi}$-dependent transformation of $ \xi$ in the first exponential.

It follows from the results of section 3 that the supercovariant derivatives of the superfields $\Xi^{\a}_i (x, \theta_i, \bar{\theta}^i)$ constructed from the Goldstinos $\xi^{\a}_i (x)$ are expressible in terms of the $\Xi^{\a}_i $  only. We do not present details, but the result is
\bea
{\cal D}_{\b}^{j} \, \Xi^{\a}_i &=& \d_{\b}{}^{\a} \d^{j}{}_i - \frac13 {\cal A}_{\b}^j{}_i^{\a} + \frac{1}{30} {\cal B}_{\b}^j{}_i^{\a} - \frac{1}{18} {\cal A}_{\b}^j{}_k^{\g} {\cal A}_{\g}^k{}_i^{\a}, \\
\bar{{\cal D}}_{\bd}^j \, \Xi^{\a}_i &=& - 2 \ri \, \left( 1 + \frac{\ri \k}{6} (\Xi.S.\Xi) \right) \, \Xi^{\b j} {\cal D}_{\b \bd} \Xi^{\a}_i,
\eea
where
\bea
{\cal A}_{\a}^i{}_j^{\b} &=& - \ri \k \left( S^i{}_k \Xi^{\b k} \Xi_{\a j} + S^i{}_k (\Xi.\Xi)^k{}_j \d_{\a}{}^{\b} + \Xi_{\a}^i \Xi _k^{\b} S^k{}_j \right)
\eea
and
\bea
\cB_{\a}^i{}_j^{\b} &=& - \kappa^2 [ -\Xi^{\b}_k S^{ik} (\Xi.S.\Xi)_{\a}{}^{\g} - \Xi^{\g}_k S^{ik} (\Xi.S.\Xi)_{\a}{}^{\b} \\
&+& S^{ik} (\Xi.\Xi)_{kl} \Xi^{\g}_m S^{lm} \delta_{\a}{}^{\b} + S^{ik} (\Xi.\Xi)_{kl} \Xi^{\b}_m S^{lm} \delta_{\a}{}^{\g} \\
&+& 2 \,
\Xi_{\a}^i (\Xi.S^2.\Xi)^{\b \g} ] \,\Xi_{\g j} \\
&+& \k^2 [ 2 (\Xi.\Xi)_{kl} S^{ik} \Xi_{\a}^l - \Xi_{\a}^i (\Xi.S.\Xi) ] \Xi^{\b}_m S^{m}{}_j
\eea
with
\be 
(\Xi.S.\Xi) = \Xi^{\a}_ i S^i{}_j \Xi^j_{\a}, \quad (\Xi.S^2.\Xi)^{\a \b}  = \Xi^{\a}_ i S^i{}_j S^j{}_k \Xi^{\b k}.
\ee
Again, it is possible to check explicitly that this algebra of covariant derivatives closes, for example,
\bea
\{ \bar{\cD}_{\ad}^i , \bar{\cD}_{\bd}^j \} \, \Xi^{\a}_k &=& - \ri \, | \kappa |^2 \epsilon_{\ad \bd} \epsilon^{ij} S_{kl} \, \Xi^{\a l}, \\
\, [ \cD_{\a \ad}, \bar{\cD}_{\bd i} ] &=& - \frac12 \, \bar{\kappa} \, \epsilon_{\ad \bd} S_{ij} \cD_{\a}^j.
\eea
 
 As in the $\cN=1$ case, the Goldstino superfields can be used  to construct composite superfields  satisfying interesting constraints. For example, it is highly nontrivial to verify that 
\be
L_{ij} = \left( 1 - \frac{\ri \k}{6} (\Xi.S.\Xi) \right) \, \Xi^{\a}_i \, \Xi_{\a j}
\ee
satisfies the constraint 
\be
\cD_{\a (i} L_{jk)} = 0.
\ee
It should be noted that this superfield does not satisfy the reality constraints  of the standard  $\cN = 2$ tensor multiplet.
\\

\noindent
{\bf Acknowledgements} \\
I am grateful to  Sergei Kuzenko for suggesting the investigation of construction of Goldstino superfieds in $AdS_4,$ for discussions during the course of the work, and for comments on the manuscript.

\begin{footnotesize}

\end{footnotesize}

\end{document}